\documentclass[11pt,a4paper]{article}
\usepackage[utf8]{inputenc}
\usepackage[left=1in,right=1in,top=1in,bottom=1in]{geometry}
\usepackage{amsmath}
\usepackage{amsthm}
\usepackage{amsfonts}
\usepackage{amssymb}
\usepackage{mathtools}
\usepackage{cite}
\usepackage{float}
\usepackage{graphicx}
\usepackage{ctable}
\usepackage{authblk}
\usepackage{makecell}
\usepackage{fancyhdr}
\usepackage[section]{placeins}
\begin{document}
\title{Low-Dimensional Manifold of Actin Polymerization Dynamics}
\author[1,2,3]{Carlos Floyd}
\author[1,2,3,*]{Christopher Jarzynski}
\author[1,2,3,$\dag$]{Garegin Papoian}
\affil[1]{Institute for Physical Science and Technology, University of Maryland, College Park, MD 20742 USA}
\affil[2]{Department of Chemistry and Biochemistry, University of Maryland, College Park, MD 20742 USA}
\affil[3]{Biophysics Program, University of Maryland, College Park, MD 20742 USA}
\affil[*]{email: gpapoian@umd.edu}
\affil[$\dag$]{email: cjarzyns@umd.edu}
\date{June 2017}

\begin{titlepage}
\maketitle
\abstract{Actin filaments are critical components of the eukaryotic cytoskeleton, playing important roles in a number of cellular functions, such as cell migration, organelle transport, and mechanosensation.  They are helical polymers with a well-defined polarity, composed of globular monomers that bind nucleotides in one of three hydrolysis states (ATP, ADP-Pi, or ADP).  Mean-field models of the dynamics of actin polymerization have succeeded in, among other things, determining the nucleotide profile of an average filament and resolving the mechanisms of accessory proteins, however these models require numerical solution of a high-dimensional system of nonlinear ODE's.  By truncating a set of recursion equations, the Brooks-Carlsson model reduces dimensionality to 11, but it remains nonlinear and does not admit an analytical solution, hence, significantly hindering understanding of its resulting dynamics. In this work, by taking advantage of the fast timescales of the hydrolysis states of the filament tips, we propose two model reduction schemes that achieve low dimensionality and linearity.  We provide an exact solution of the resulting linear equations and use it to shed light on the dynamical behaviors of the full BC model, highlighting the relative ordering of the timescales of various collective processes, and explaining some unusual dependence of the steady-state behavior on initial conditions.}
\end{titlepage}

\section{Introduction}
Actin filaments are an integral part of the cytoskeleton and are involved in functions such as controlling cell shape, cell motility, organelle redistribution, and mechanical coupling with the cellular environment.  These filaments are formed of globular monomers which polymerize in a nonequilibrium process that $\textit{in vivo}$ is modulated by an array of accessory proteins.  They are helical and polar, with distinct plus (``barbed") and minus (``pointed") ends at which monomers have different rates of association and dissocation \cite{lodish1995molecular}.  Hydrolysis of cellular ATP leads to filament ``treadmilling" which drives the polymerization process away from equilibrium and allows actin networks to be responsive to different cellular cues \cite{theriot1991actin, schmidt1998signaling, vogel2006local}.  Each actin monomer molecule is bound to a nucleotide, which is in one of several hydrolysis states: adenosine triphophate (ATP), adenosine diphosphate (ADP), or an intermediate state ADP-Pi, in which ADP is still bound to a hydrolyzed inorganic phosphate molecule.  Release of inorganic phosphate by ADP-Pi converts it to ADP \cite{carlier1997actin}.  The hydrolysis state of the bound nucleotide has dramatic effects on the kinetic polymerization and depolymerization rate constants of the globular monomer  \cite{pollard1986rate}.  In addition, the hydrolysis states of the filament monomers affect the binding affinity of accessory proteins and structural properties such as persistence length \cite{andrianantoandro2006mechanism, isambert1995flexibility}.  Thus it is of interest to be able to predict the hydrolysis state of the nucleotide bound to each actin monomer in a filament, or at least the fraction of actin monomers bound to nucleotides in a certain hydrolysis state.  

Over several decades a variety of models describing actin polymerization dynamics have been put forward, and these models have evolved alongside the growth of experimental knowledge about the nature of actin.  Some early models tracked the number of filaments with a certain degree of polymerization under different assumptions about the filament polarity, geometry, and the cooperativity of polymerization, among other factors \cite{wegner1975kinetics, cooke1975role, wegner1976head, pantaloni1985model,  hill1982subunit, korn1987actin}.  Polymerization and depolymerization rate constants for ATP and ADP-bound actin were measured for the first time in 1986 \cite{pollard1986rate}.  A subset of more recent models have investigated aspects such as the effects of accessory proteins on actin polymerization via tracking the time-varying concentration of actin monomers distinguished by their polymerization state and by the hydrolysis state of the nucleotide they are bound to.  A variable is assigned to the concentration of each species and complexes between certain species, and equations of motion in terms of mean-field mass-action kinetic rate constants are written for each.  The resulting coupled ordinary differential equations (ODEs) are solved numerically, and the effects of varying parameters such as reservoir ATP/ADP disequilibrium, total filament concentration, fraction of capped plus ends, free actin concentration, and profilin concentration are investigated \cite{dufort1996profilin, bindschadler2004mechanistic}.  One point of contention is whether transitions between hydrolysis states of polymerized monomers occurs in a random fashion, in which hydrolysis states of a monomer's neighbors do not affect the hydrolysis rate of that monomer, or in a vectorial fashion, in which an ATP-bound monomer will only hydrolyze ATP if its neighbor towards the minus end is ADP-Pi bound, leading to a contiguous ATP-bound cap at the plus end.  Recent models suggests that the truth is in the middle, such that coupling exists in ATP cleavage rates between neighboring polymerized monomers, but not such that the process is truly vectorial \cite{li2009actin, stukalin2006atp}.  Most mean-field models make the assumption of random ATP hydrolysis for simplicity.  

An important disadvantage of such mean-field models aimed at resolving the roles of accessory proteins is that their level of detail inhibits analytical solutions to the time courses and, hence, obscures deeper insights into dynamical behaviors of these systems.  While this approach has successfully allowed modelers to, for example, rule out certain mechanisms of profilin's action on critical concentrations \cite{yarmola2008effect}, one might ask what is the simplest such model that reproduces time courses from more detailed models.  This is the aim of the present work.  The model reduction here is based on a 2009 model by Brooks and Carlsson (BC) \cite{brooks2009nonequilibrium}, which presents a system of differential equations that admits only numerical solution but does not include extra detail by accounting for accessory proteins.  It is useful for predicting the process of polymerization when a pool of monomers are added to an initial concentration of seed filaments, and is sufficiently simple to be incorporated into larger-scale cellular models without too much additional computational cost. 

In this work, we report on two successive reduction schemes of the 11-dimensional BC model: a quasi-steady-state approximation that leverages fast dynamics of the filament tips, leading to a 5-dimensional system of ODEs, and a subsequent linearization approximation. The latter equations admits an analytical solution whose implications we investigate, revealing interesting features of actin polymerization process projected on the slow dynamical manifold.  Our analytical model reduction approaches show excellent agreement with the results obtained from stochastic simulations of the full BC model and also when compared with diffusion mapping analyses of stochastic trajectories.

\section{Methods}
\subsection{Brooks-Carlsson Model}
\begin{figure}[H]
\centering
\includegraphics[width=15 cm]{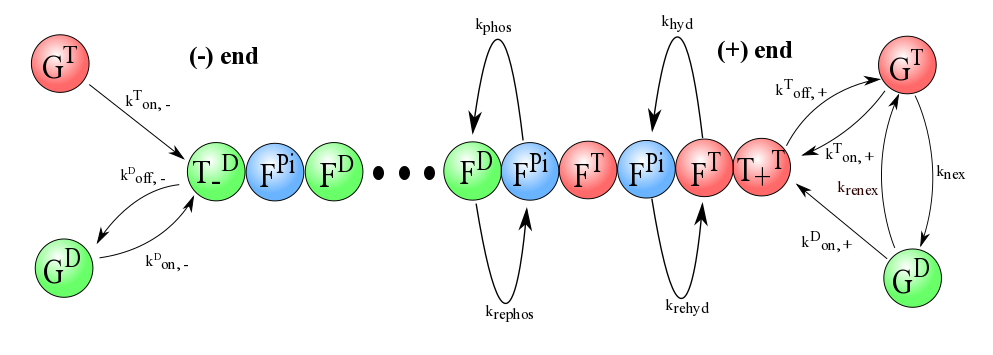}
\caption{A linear actin filament, with monomers colored according to hydrolysis state.  Random, as opposed to vectorial, hydrolysis is assumed here.  Some of the reactions are slightly misleading as drawn: for example a hydrolysis reaction converting $F^T$ to $F^{Pi}$ would happen at a single location in the filament, i.e. the monomer would not change into its neighbor as shown here.  Also, the polymerization of $G^T$ onto the minus end would convert $T^D_-$ into $T^T_-$, and a similar statement applies to $G^D$ polymerizing to the plus end.  The depolymerization of $T^{Pi}_\pm$ is not shown.  }
\label{fig1}
\end{figure}

The BC model of actin polymerization is an 11-dimensional system of ordinary differential equations tracking the concentration of non-tip actin monomers in different states as well as the concentrations of filament tip monomers in different states \cite{brooks2009nonequilibrium}.  It is assumed that the number concentration of filaments $N$ remains constant, implying an absence of filament nucleation, splitting, or joining.  Additionally, the total concentration of actin monomers $M$ is assumed to remain constant, such that actin monomers are not created or destroyed in any reaction.  Since there are typically many actin monomers belonging to a given actin filament, we have $N \ll M$.  All species are assumed to be well-mixed and in large enough quantities to be treated effectively via a mean field description.  In other words, the size of the stochastic fluctuations is negligible compared to the concentrations of the species.  Unpolymerized (globular) actin monomers are referred to as G-actin, while polymerized (filamentous) actin monomers are referred to as F-actin.  Actin filaments are helical, but they are more easily modeled as linear chains, which is a realistic approximation if one assumes that the reaction propensities of a given F-actin monomer is determined only by the nucleotide bound by that monomer and not by the monomer's neighbors.  Such a chain is displayed in Figure \ref{fig1}, along with some of the reactions allowing interconversion between monomer types.  The variables representing the concentrations of these actin species are superscripted by the hydrolysis state of the bound nucleotide (for what follows we refer more simply to the monomer being in a certain hydrolysis state as opposed to the nucleotide attached to a monomer as being in that state). The hydrolysis states are ATP, ADP-Pi, and ADP, denoted T, Pi, and D, respectively.  The tip monomers are denoted $T$ and are further subscripted according to which tip they are on.  Thus, for example, the concentration of tip monomers at the plus end bound to ADP-Pi is denoted $T_+^{Pi}$.  Because inorganic phosphate rapidly dissociates from G-actin, $G^{Pi}$ is taken to be 0 and is not tracked.  With the 3 hydrolysis states of each of the 2 tips, the 3 states of the F-actin and the 2 states of G-actin, the tracked variables are 11 in number.  

The different monomer types interconvert through chemical reactions.  These reactions can be roughly classified as polymerization/depolymerization reactions which change G actin to F actin and vice versa, or as reactions in which the ATP hydrolysis state of the monomer changes.  The evolution of the concentrations of the different tip monomer hydrolysis states in principle depends on the hydrolysis state of the monomer adjacent to the tip, which itself depends on the hydrolysis state of the next monomer in the filament, and so on.  Every monomer in the filament then requires keeping track of, causing the dimensionality of the model to be roughly equal to the degree of polymerization of a filament, which typically contains hundreds of monomers.  The major accomplishment of the BC model is to truncate this set of recursion equations by assuming that the hydrolysis state of the monomer adjacent to the tip depends only on the hydrolysis state of the tip monomer.  Using the results of stochastic simulations of a more complete model, they write empirical equations to capture these relationships, and in doing so they close off a 11-dimensional subset of of the original hundreds of equations.  They find close agreement between their truncated model and the full stochastic simulation over a wide and realistic range of parameters.  The equations of motion for the 11 variables in the BC model can be written as a nonlinear system of ODE's:
\begin{equation}
\dot{\bold{x}} = \bold{f}(\bold{x})
\label{eq1}
\end{equation}
where $\bold{x}$ is a vector of the concentrations of the 11 species, and $\bold{f}$ is a nonlinear vector-valued function of $\bold{x}$.  This system of equations is solvable only numerically, but the results match well with simulations which accurately model experimental data \cite{brooks2008actin}.   The steady-state vector $\bold{x}^\text{ss}$ satisfying $\bold{f}(\bold{x}^\text{ss}) = \bold{0}$ is unique for a given value of $N$ and $M$ and under the condition that all concentrations be real and non-negative, and it is attracting.  We give more details of the BC model in Appendix A, where we list the 11 ODE's.  

A separation of timescales exists between the dynamics the monomer states and those of tip states, the latter of which evolve much more rapidly than the former.  Thus we partition the vector $\bold{x}$ into slow and fast variables: $\bold{x}_\text{s} \equiv (G^T, \ G^D, \  F^T, \  F^{Pi}, \ F^D)^\intercal$, $\bold{x}_\text{f} \equiv (T^T_+, \ T^{Pi}_+, \ T^D_+, \ T^T_-, \ T^{Pi}_-, \ T^D_-)^\intercal$, where the subscripts s and f refer to ``slow" and ``fast".  Terms of comparable magnitude appear in the equations of motion for both $\bold{x}_\text{f}$ and $\bold{x}_\text{s}$, but the elements in $\bold{x}_\text{f}$ are typically much smaller than those in $\bold{x}_\text{s}$ because $N \ll M$, and so $\bold{x}_\text{f}$ experiences greater acceleration for a given forcing term, and its dynamics are therefore faster.  With these new variables, Equation \ref{eq1} can be usefully rewritten as follows:

\begin{eqnarray}
\label{eq2}
&&\dot{\bold{x}}_\text{s}= \bold{A}\bold{x}_\text{s} + \bold{B}\bold{x}_\text{f}\\  
\label{eq3}
&&\dot{\bold{x}}_\text{f} = \bold{h}(\bold{x}_\text{s}, \ \bold{x}_\text{f}) 
\end{eqnarray} 
where $\bold{A}$ and $\bold{B}$ are matrices whose off-diagonal elements are combinations of kinetic rate constants and whose columns sum to zero due to conservation of actin, and $\bold{h}$ is a nonlinear vector-valued function containing terms that are up to cubic products of variables.   

The BC model is a major advancement towards a more computationally accessible model of the dynamics of actin polymerization, however we might ask for several other features in such a model.  We ask that it: (1) be low-dimensional, (2) capture the interesting and important timescales, and (3) be exactly solvable.  To meet these goals, we make the decision to only track the vector $\bold{x}_\text{s}$.  If we were to track the concentration of the tip monomer states, i.e. the elements in $\bold{x}_\text{f}$, we would automatically increase the dimensionality of the model, and we will discuss reasons why we may assume that the tip monomers are evolving in such a way that keeping track of them explicitly is unnecessary.  We make two approximations for how to treat $\bold{x}_\text{f}$ in Equation \ref{eq2}, first utilizing the fact that a separation of timescale exists between the dynamics of $\bold{x}_\text{f}$ and $\bold{x}_\text{s}$, and then utilizing the fact that $ | \bold{B}\bold{x}_\text{f}| \ll |\bold{A}\bold{x}_\text{s}|$ except the system is near near steady-state, at which point these terms have comparable magnitudes.  This second fact arises since $\bold{x}_\text{f}$ contains terms up to $\mathcal{O}(N)$ and $\bold{x}_\text{s}$ contains terms up to $\mathcal{O}(M)$.  

It is also possible to demonstrate that a low-dimensional description of the slow dynamics is a valid approximation through the use of diffusion mapping on a stochastically generated data set based on the BC model.  We describe this analysis in Supporting Information 1.  

\subsection{Quasi-Steady-State Approximation}

The quasi-steady-state approximation (QSSA) relies on the assumption that certain variables have much faster dynamics than other slower variables.  This separation of timescales allows one to assume that the fast variables $\bold{x}_\text{f}$ are always in equilibrium with respect to the slow variables $\bold{x}_\text{s}$, and therefore that the values of the slow variables determine the values of the fast variables at any moment.  We can imagine that $\bold{x}_\text{f}$ is effectively being ``dragged around" by the values of the elements in $\bold{x}_\text{s}$.  So, we can solve for functions $\bold{x}_\text{f}^\text{eq}(\bold{x}_\text{s})$ relating the quasi-equilibrated fast variables in terms of the slow variables by imagining holding $\bold{x}_\text{s}$ fixed and finding the equilibrium values of $\bold{x}_\text{f}$.  This amounts to the condition $\bold{h}(\bold{x}_\text{f}^\text{eq} \ ; \ \bold{x}_\text{s})=\bold{0}$.  The functions $\bold{x}_\text{f}^\text{eq}(\bold{x}_\text{s})$ are then substituted in the equations of motion for the slow variables giving the closed system of equations
\begin{equation}
\label{eq4}
 \dot{\bold{x}}_\text{s} = \bold{A}\bold{x}_\text{s} + \bold{B}\bold{x}_\text{f}^\text{eq}(\bold{x}_\text{s})
\end{equation}
This subsystem is lower-dimensional, though it is nonlinear since $\bold{x}_\text{f}^\text{eq}(\bold{x}_\text{s})$ is nonlinear, and it describes the evolution of the system on the slow timescales.

\begin{figure}[H]
\centering
\includegraphics[width=15 cm]{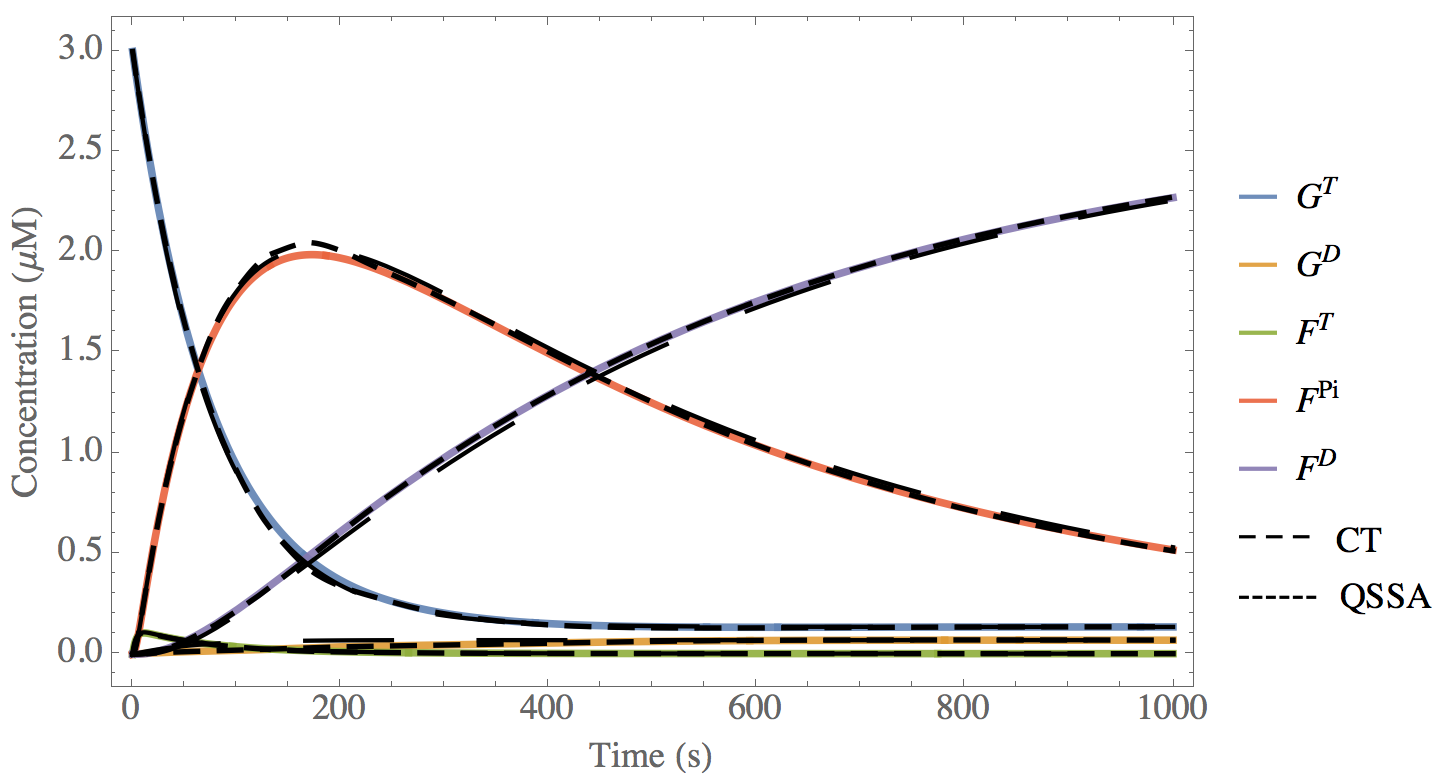}
\caption{Time course of the concentrations of the various species in $\bold{x}_\text{s}$ following the addition of 3 $\mu \text{M}$ of $G^T$ actin to a bath with a number concentration $N=1$ nM of seed filaments.  This image displays the numerical integration of the BC model as solid colors, the numerical integration of the QSSA model (Section 2.2) as a short dashed line, and the exact solution of the CT model (Section 2.3) as a long dashed line.}
\label{fig2}
\end{figure}

In the BC model, the condition $\bold{h}(\bold{x}_\text{f}^\text{eq} \ ; \ \bold{x}_\text{s})=\bold{0}$ implies the following algebraic systems of equations (see Appendix A):
\begin{eqnarray}
\label{eq5}
&&\frac{dT^T_\pm}{dt} = 0 \notag \\
&&\frac{dT^{Pi}_\pm}{dt} = 0 \\
&&\frac{dT^D_\pm}{dt}=0 \notag 
\end{eqnarray}
Only four of these six equations are linearly independent due to the conservation of number of plus and minus end filament tips, so we use the following supplementary equations to find a solution of the combined systems of equations:
\begin{equation}
\label{eq6}
T^T_\pm + T^{Pi}_\pm + T^D_\pm = N
\end{equation} 
System of Equations \ref{eq5}, \ref{eq6} can be solved numerically resulting in tabulated functions of the forms $T^T_\pm(G^T, \ G^D)$, $T^{Pi}_\pm(G^T, \ G^D)$, and $T^D_\pm(G^T, \ G^D)$.  These functions do not depend on $F^T$, $F^{Pi}$, and $F^D$ because these variables do not enter into the the function $\bold{h}$.  Subsequently, the nonlinear (slow) system described by Equation \ref{eq4} is then numerically integrated.  Figure \ref{fig2} displays a comparison of the QSSA approximation to the original 11-dimensional BC model.  

This approximation succeeds in reducing the dimensionality of the model to 5.  We note, however, that the dynamics of this model lie on a 4-dimensional submanifold of the full 5-dimensional manifold due to the fact that $\bold{A}$ and $\bold{B}$ are both singular, corresponding to conservation of actin.  This model also captures the interesting timescales corresponding to polymerization events and dynamics of the hydrolysis states of F and G actin, while it neglects the fast dynamics corresponding to such events taking place at the tips.  However the model is still highly nonlinear and analytically unsolvable, so we propose an alternative model reduction scheme.  

\subsection{Constant Tip Approximation} 

In addition to having sufficiently fast dynamics as to be effectively described as in quasi-equilibrium with respect to $\bold{x}_\text{s}$, the vector $\bold{x}_\text{f}$ is also small in magnitude compared to $\bold{x}_\text{s}$, and this can be utilized to do drastically simplify the QSSA model.  Although the tip dynamics are fast, one can profitably assume that the concentration of filament tip states, that is, the elements in $\bold{x}_\text{f}$, are constant in time.  We refer to this assumption as the constant tip (CT) approximation.  Certainly this assumption is not realistic since the tips have fast dynamics, but the effect of this error on the equations of motion of the other actin species is small in the regime where the number concentration of filaments is much smaller than the total amount of actin in the system, i.e. when $N \ll M$, and when the tips states quickly attain their steady-state values.  This assumption allows the model to be reduced to a 5-dimensional linear system of equations which can be solved analytically.  The procedure is to replace the dynamical variables $T^i_j$ by constants $N \Gamma^i_j$, chosen such that the steady-state of the CT model coincides with the steady-state of the BC model.  We define $\Gamma^i_j$ as the fraction of the filament tips at the $j$ (plus or minus) end that are in the $i$ (T, Pi, D) hydrolysis state.  These constants determine the rate of depolymerization reactions, and replacing the variables  $T^i_j$ with them causes the term $\bold{B}\bold{x}_\text{f}$ to be a constant source and sink term in Equation \ref{eq2}, which as a result becomes linear.  

The steady-states of the BC and CT models will be the same if
\begin{eqnarray}
\label{eq7}
&&N \Gamma^T_\pm = \lim_{t \rightarrow \infty} T^T_\pm(t) \notag \\
&&N \Gamma^{Pi}_\pm = \lim_{t \rightarrow \infty} T^{Pi}_\pm(t) \\
&&N \Gamma^D_\pm = \lim_{t \rightarrow \infty} T^D_\pm(t) \notag 
\end{eqnarray}
In other words, the constant tip state fractions in the CT model should be chosen as the steady-state values of the tip state fractions in the BC model.  As a result, only the approach to steady-state will be different between the two models.  These limiting values can be found by numerically solving the algebraic system of equations 

\begin{eqnarray}
\label{eq8}
&& \bold{f}(\bold{x}^\text{ss}) = 0 \notag \\
&& T^T_\pm + T^{Pi}_\pm + T^D_\pm = N \\
&& G^T + G^D + F^T +F^{Pi} + F^D = M \notag
\end{eqnarray}
where $\bold{x}^\text{ss}$ is the 11-dimensional steady-state vector of concentrations in the BC model, and taking the real non-negative solution.  Equation \ref{eq8} determines the values of $\Gamma^i_j$ which will be unique for a given $N$ and $M$.  

We define 
\begin{equation}
\label{eq9}
\bold{b} \equiv \bold{B}\bold{x}_\text{f}^\text{ss}
\end{equation}
where $\bold{x}_\text{f}^\text{ss}$ contains the constants $N \Gamma^i_j$ instead of the variables $T^i_j$.  The equation of motion for $\bold{x}_\text{s}$ is 

\begin{equation}
\label{eq10}
\dot{\bold{x}}_\text{s} = \bold{A}\bold{x}_\text{s} + \bold{b}
\end{equation}
where 
\begin{eqnarray}
\label{eq11}
&& \bold{A}=\begin{pmatrix}
-a -b & c & 0 & 0 & 0\\
b & -c -d & 0 & 0 & 0\\
a & 0 & -e & f & 0\\
0 & 0 & e & -f-g & h\\
0 & d & 0 & g & -h
\end{pmatrix} \\
\label{eq12}
&& \bold{b} = \begin{pmatrix}
i\\
j+k\\
-i\\
-k\\
-j
\end{pmatrix}
\end{eqnarray}
with 

\begin{equation}
\label{eq13}
\begin{matrix*}[l]
a = N(k_\text{on, +}^T + k_\text{off, +}^T) && g = k_\text{phos}\\
b = k_\text{nex}  &&  h = k_\text{rephos}\\
 c = k_\text{renex}&&   i = N(\Gamma_+^T k_\text{off, +}^T + \Gamma_-^T k_\text{off, -}^T) \\
d = N(k_\text{on, +}^D + k_\text{off, +}^D)  && j = N(\Gamma_+^D k_\text{off, +}^D + \Gamma_-^D k_\text{off, -}^D)\\ 
e = k_\text{hyd} && k = N(\Gamma_+^{Pi} k_\text{off, +}^{Pi} + \Gamma_-^{Pi} k_\text{off, -}^{Pi}) \\
f = k_\text{rehyd}
\end{matrix*}
\end{equation}

Equation \ref{eq10} is a linear system of differential equations which can be solved exactly.  One point of difficulty in solving this system is that the columns of $\bold{A}$ sum to zero due to conservation of actin, causing $\bold{A}$ to be singular.  In chemical reaction network theory, one often has such systems with linear conservation laws.  If the system is linear, with only first order or pseudo-first order reactions, then such a singular system can be solved cleanly by a method using the Drazin inverse $\bold{A}^\mathcal{D}$ of the matrix $\bold{A}$.  We believe that this method has certain advantages over other approaches to solving singular systems of differential equations.  If $\bold{A}$ has Jordan decomposition 

\begin{equation}
\label{eq14}
\bold{A} = \bold{V}
\begin{pmatrix}
J_1 & 0\\
0 & J_0\\
\end{pmatrix}
\bold{V}^{-1} 
\end{equation}
where $J_1$ and $J_0$ correspond to the non-zero and zero eigenvalues respectively, then

\begin{equation}
\label{eq15}
\bold{A}^\mathcal{D} = \bold{V}
\begin{pmatrix}
J_1^{-1} & 0\\
0 & 0\\
\end{pmatrix}
\bold{V}^{-1}
\end{equation}
In Supporting Information 2 we go over the details of solving Equation \ref{eq10} as well as the advantages of this method \cite{campbell1976applications}.  The solution is 
\begin{equation}
\label{eq16}
\bold{x}_\text{s} = \bigg(-\bold{A}^\mathcal{D} + \bigg(\bold{I}-\bold{A}\bold{A}^\mathcal{D}\bigg)t\bigg[\sum_{n=0}^{k-1}\frac{\bold{A}^nt^n}{(n+1)!}\bigg] + e^{\bold{A}t}\bold{G}\bigg)\bold{b} = \bigg(-\bold{A}^\mathcal{D} +  e^{\bold{A}t}\bold{G}\bigg)\bold{b}
\end{equation}  
where in the last step we have simplified using properties of $\bold{A}$ for this system.  The matrix 
\begin{equation}
\label{eq17}
\bold{G} = \frac{1}{|\bold{b}|^2}\bold{x}_\text{s}(0)\bold{b}^\intercal + \bold{A}^\mathcal{D}
\end{equation}
encodes information about the initial conditions $\bold{x}_\text{s}(0)$.

\section{Results}

\subsection{Eigenvalues of $\bold{A}$}
Perhaps the most important benefit of having an exactly solvable model is the ability to formally determine the eigenvalues governing the linear dynamics.  The expressions for the eigenvalues will shed light on the timescales that describe the different chemical processes.  In our modeling we have included the reverses of kinetically dominant forward reactions, and we have set the rate constants of these reactions to be equal to something on the order of the corresponding forward reaction rate constants multiplied by a small parameter $\epsilon$.  Thus $b = \epsilon b^*$, where $b^* \sim c$, $f=\epsilon f^*$, where $f^* \sim e$, and $h=\epsilon h^*$, where $h^* \sim g$.   By writing reverse rate constants this way, we can Taylor expand the expressions for the eigenvalues around the point $\epsilon = 0$ to simplify the result.  Doing so to first order in $\epsilon$, we have 
\begin{align}
\label{eq18}
\lambda_1 &= 0 \\
\label{eq19}
\lambda_2 & = \frac{1}{2}\bigg(-a-\epsilon b^* -c-d + \sqrt{(a-d+\epsilon b^*-c)^2 +4\epsilon b^* c}\bigg) \notag\\
& \approx - c - d + \frac{b^* c}{a-c-d}\epsilon  \\
\lambda_3& = \frac{1}{2}\bigg(-a-\epsilon b^* -c-d - \sqrt{(a-d+\epsilon b^*-c)^2 +4\epsilon b^* c}\bigg) \notag\\
\label{eq20}
&  \approx -a - \frac{b^*(a-d)}{a-c-d}\epsilon \\
\lambda_4 &=  \frac{1}{2}\bigg(-e-\epsilon f^* -g - \epsilon h^* + \sqrt{(e-g +\epsilon f^* - \epsilon h^*)^2 +4\epsilon f^* g}\bigg) \notag \\
\label{eq21}
& \approx -g -\bigg(h^*-\frac{g f^* }{e-g}  \bigg)\epsilon \\
\lambda_5& = \frac{1}{2}\bigg(-e-\epsilon f^* -g - \epsilon h^* - \sqrt{(e-g +\epsilon f^* - \epsilon h^*)^2 +4\epsilon f^* g}\bigg) \notag \\
\label{eq22}
& \approx -e -\frac{ e f^*}{e-g}\epsilon 
\end{align}
The fact that the nonzero eigenvalues are negative implies the stability of the steady-state.  These eigenvalues have no dependence on $M$, the total concentration of actin monomers, but they do depend on $N$, since this term appears in the expressions for $a$ and $d$.  For the parameterization used here and with $N=1 \ \text{nM}$, the eigenvalues can be ordered by magnitude as follows:
\begin{equation}
\label{eq23}
|\lambda_5| > |\lambda_3| > |\lambda_2| > |\lambda_4|>|\lambda_1|
\end{equation}
  
Equations \ref{eq18}-\ref{eq22} are in terms of reaction rate constants and can be interpreted as representing certain collective subprocesses in the chemical system corresponding to combinations of those reactions.  The ordering indicates the comparative rates of those subprocesses.  We interpret these the zeroth order terms of the eigenvalues as follows:
\begin{itemize}
\item $\lambda_1$ being equal to zero arises from the fact that actin is conserved in this system, causing $\bold{A}$ to be singular and its rank less than its row number.  Equivalently, one can say that the dynamics unfold on a 4-dimensional submanifold of the 5-dimensional manifold, and that this submanifold is determined by $M$. 
\item $\lambda_2$ represents the combination of the forward nucleotide exchange reaction ($G^D \rightarrow G^T$) and the polymerization of $G^D$.  Both of these reactions convert $G^D$ into another species, so this eigenvalue represents the subprocess of $G^D$ depletion.   
\item $\lambda_3$ represents the polymerization of $G^T$.  
\item $\lambda_4$ represents the release of phosphate by $F^{Pi}$ to form $F^D$.    
\item $\lambda_5$ represents  the hydrolysis of ATP converting $F^T$ to $F^{Pi}$. 
\end{itemize}
Whether the magnitudes of these eigenvalues are increased or decreased due to the presence of reversible reactions (i.e. when $\epsilon > 0$) depends on the parameterization, since the sign of of the first order terms depend on the comparative sizes of certain parameters.

In the full BC 11-dimensional model, one can numerically evaluate the Jacobian matrix of $\bold{f}(\bold{x})$ at steady-state:
\begin{equation}
\label{eq24}
\bold{J^*} \equiv \frac{\partial \bold{f}}{\partial \bold{x}} \bigg | _{\bold{x} = \bold{x}^\text{ss}}
\end{equation}
We find that, for the same parameterization, the smallest four non-zero eigenvalues of $\bold{J^*}$ are almost exactly equal to the non-zero eigenvalues of $\bold{A}$.  This implies that the CT model has captured the slowest dynamics of the BC model by ignoring the processes involving the tip monomers.  The main benefit of the constant tip approximation is that these slow linear dynamics can now be easily analyzed.  These dynamics provide information about the polymerization process, the nucleotide composition of the filaments, and nucleotide exchange of globular actin.  For most purposes, these aspects are of primary interest, and the processes at the tips are of lesser importance.  

\subsection{Steady-State Concentrations}
One might expect that if we increase the amount of actin in the system by a different choice of initial conditions, the concentrations of the different species at steady-state would change.  An interesting feature of this system is that this is true only for some species, and which species it is true for depends on whether or not we have included reversible reactions (if $\epsilon > 0$).  Additionally, this can be shown to be true in both the CT model and the BC model.  We demonstrate it first in the CT model.  

We find the steady-state vector of concentrations $\bold{x}_\text{s}^\text{ss}$ by taking the limit of Equation \ref{eq16} as $t \rightarrow \infty$:

\begin{equation}
\label{eq25}
 \bold{x}_\text{s}^\text{ss} \equiv \lim_{t \to\infty} \bold{x}_\text{s}(t)  = -\bold{A^\mathcal{D}}\bold{b}+\bold{C}\bold{G}\bold{b}
\end{equation}
where 
\begin{equation}
\label{eq26}
\bold{C} \equiv \lim_{t \to \infty} e^{\bold{A}t} 
\end{equation}
We diagonalize $\bold{A}$ as $\bold{U}\bold{D}\bold{V^\dagger}$ and use it in the expression for $\bold{C}$:
\begin{equation}
\label{eq27}
\bold{C} = \bold{U} \lim_{t \to \infty} e^{\bold{D}t}\bold{V^\dagger}
\end{equation}
With the exception of $\lambda_1$, which is zero, the eigenvalues of $\bold{A}$ are negative, so we have

\begin{align}
\label{eq28}
\bold{C}&=\bold{U} \ \text{diag}(1,0,0,0,0) \ \bold{V^\dagger} \notag\\
 &= \begin{pmatrix}
0 & 0 & 0 & 0 & 0 \\
0 & 0 & 0 & 0 & 0 \\
\frac{f h}{e g} & 0 & 0 & 0 & 0 \\
\frac{h}{g} & 0 & 0 & 0 & 0 \\
1 & 0 & 0 & 0 & 0 \\
\end{pmatrix} \bold{V^\dagger}
\end{align}
where $(0,0,\frac{f h}{e g},\frac{h}{g},1)^\intercal$ is the eigenvector corresponding to $\lambda_1$.  Thus the top two rows of $\bold{C}$ are zero, and the top four rows would be zero if we exclude reversible reactions.  This is also true of the term $\bold{C}\bold{G}\bold{b}$.  Now, the matrix $\bold{G}$ is the only place where the initial conditions appear in Equation \ref{eq25}.  So if the top two rows $\bold{C}\bold{G}\bold{b}$ are zero, then the first two elements of $\bold{x}_\text{s}^\text{ss}$ cannot have any dependence on initial conditions.  In other words, $G^D$ and $G^T$ always reach the same concentrations at steady-state no matter what the initial concentrations of any of the species are.  If we have no reversible reactions, the same is also true for the species $F^T$ and $F^{Pi}$.  

Consider the following thought experiment, assuming for simplicity $\epsilon = 0$. We start with some initial amount $M$ of actin in any form and let the system come to steady-state.  We then add an amount $\Delta M$ more actin to the system, in any form, and wait until the system has reached steady-state again.  We would find that the only difference between the two steady-states is that the concentration of $F^D$ had increased by $\Delta M$.  If $\epsilon >0$, then we would find that the steady-state values of $F^T$, and $F^{Pi}$, $F^D$ had all increased, and the sum of these changes would be $\Delta M$.  

This lack of dependence of the steady-state concentrations of some species on $M$ is not an artifact of the constant tip approximation; it is also the case in the BC model.  It can not be shown to be true by taking the limit $t \rightarrow \infty$ as is the case here, but it can instead be shown by observing that a subset of the system of algebraic equations $\bold{f}(\bold{x}^\text{ss})=0$ is closed, and that a solution for the subset could be obtained without specifying $M$.  This implies that the steady-state concentrations of the species represented by that solution have no dependence on $M$.  We give the details of this argument in Appendix B.  

\section{Discussion}
We have argued that the dynamics of the polymerization of actin monomers into filaments can effectively be subdivided as follows: fast nonlinear dynamics govern the states of the filament tip monomers, and slow linear dynamics govern the change in polymerization and hydrolysis states of non-tip monomers.  One cannot completely separate the tip monomer dynamics from the non-tip monomer dynamics because they are highly coupled; the tip monomer states depend on the concentration of $G^T$ and $G^D$ through polymerization reactions, and the non-tip monomer states depend on the tip monomer states via depolymerization reactions.  Because of the typical size of $N$ compared to $M$, the non-tip monomer states depend only comparatively weakly on the tip monomer states during most of a typical trajectory.  We have shown two ways to approximate this coupling to achieve a significant reduction in dimensionality of the system.  First, in the QSSA model, it is assumed that, on the slow timescale, the number of tips in a certain hydrolysis state depends only on $G^T$ and $G^D$.  This allows one to write a closed system of equations for the equations of motion of the non-tip monomers, describing evolution of the entire system on the slow 5-dimensional submanifold of the full 11-dimensional space.  This model is physically realistic and highly accurate because it preserves the dependence of the tip monomer states on the concentration of the non-tip monomers, however the resulting equations of motion are analytically intractable and only integrable numerically.  

In the CT model, we make the seemingly unrealistic assumption that the tip monomers have no dependence on the non-tip monomers and in fact do not evolve at all, but remain fixed for all times at their steady-state values.  In this sense we turn the tip monomer concentrations from variables into constants, and the equations of motion for the non-tip monomers becomes 5-dimensional and linear with the depolymerization terms involving the tip monomers entering as a non-homogenous term $\bold{b}$.  By choosing the fixed values of the tip monomers as the resting values, we ensure that the steady-state of the two models will be the same.  The CT assumption is valid because of the weak dependence of the non-tip monomer states on the tip monomer states.  In other words, $\bold{b}$ is comparatively small, and the discrepancy between trajectories of the CT and BC model due to $\bold{b}$ not containing realistic values for all times is not pronounced.  In exchange for the cost of this error, there is an important benefit, which is the ability to write symbolic expressions for the eigenvalues of the matrix $\bold{A}$.  We note that these eigenvalues agree with the numerically calculated smallest nonzero eigenvalues of $\bold{J^*}$ of the full 11-dimensional model, indicating that indeed the linear non-tip monomer dynamics are the slowest of all of the processes in the BC model.  In addition, qualitative results about the nature of the dependence of the steady-state concentrations on the initial conditions agree for the full and reduced model.  We have also showed dimensionality reduction to be possible by diffusion map analysis of a simulated trajectory of the BC model (Supporting Information 1).  The results of this analysis indicate that fewer than 6 or 7 dimensions suffice to faithfully reproduce the polymerization dynamics. 

\begin{figure}[H]
\centering
\includegraphics[width=13 cm]{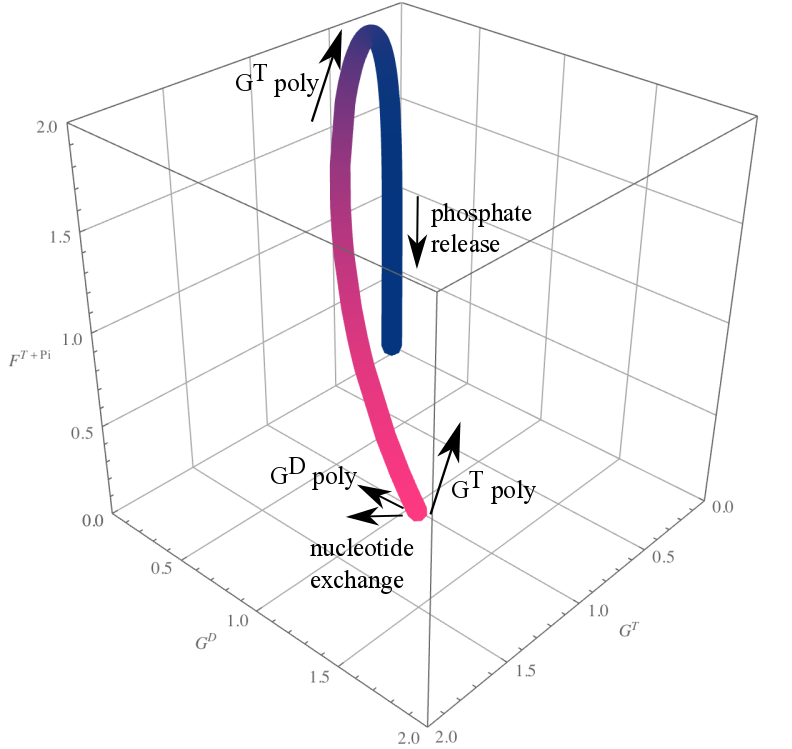}
\caption{Visualization of a 3,000 s trajectory of the CT model beginning from 1 $\mu \text{M}$ of $G^T$ and 1 $\mu \text{M}$ of $G^D$, with $N = 1 $ nM and with $F^D$ not visualized.  The curve is colored according to time, with pink representing early times.  All units are $\mu \text{M}$.  Vectors are drawn, not to scale, and labeled to represent the direction that certain reactions pull the trajectory and at which point in the trajectory those pulls become dominant.}
\label{fig3}
\end{figure}

Eigenvalue analysis of $\bold{A}$ allows one to understand the timescales that govern the linear non-tip monomer dynamics as well as how these timescales depend on the parameters.  These timescales approximately represent the following processes: depletion of $G^D$ via polymerization and conversion to $G^T$, polymerization of $G^T$, hydrolysis of ATP converting $F^T$ to $F^{Pi}$, and phosphate release converting $F^{Pi}$ to $F^D$.  As might be expected, the timescales involving polymerization depend on $N$ and their magnitude compared to that of the other timescales may change significantly.  We have treated the presence of the reverses of some nearly irreversible reactions essentially as perturbations by regarding the rate constants of backward reactions as equal to the rate constant of the corresponding forward reaction multiplied by a small parameter $\epsilon$.  As shown above, the inclusion of these reactions introduce small corrections to the eigenvalues.  These time scales allow one to understand a typical trajectory of the system.  Such a trajectory can be visualized in three dimensions by combining multiple species into a single species and choosing to not to visualize a variable whose value is determined by the other three due to conservation of actin.  In Figure \ref{fig3}, we combine $F^T$ and $F^{Pi}$ together, since they have similar structural properties in the filament.  Thus we neglect the timescale corresponding to ATP hydrolysis.  In the trajectory depicted, $G^T$ and $G^D$ are quickly made small via polymerization and nucleotide exchange reactions.  The polymerization of $G^T$ causes $F^{T+Pi}$ to increase and when $G^T$ has become small, $F^{T+Pi}$ converts to $F^D$ via the slow process of phosphate release, and at the end, nearly all of the actin is in the form $F^D$.

\section*{Conclusion}

The main results of this work are the elucidation of the degree to which neglecting tip monomer dynamics during actin polymerization is a passable assumption and the resulting insight into the hierarchy of processes involved in the slow linear dynamics of the non-tip monomers.  The CT model is overly simple but useful to understand basic features of the polymerization process.  In other more detailed models, actin related proteins are incorporated either via introducing new variables representing the proteins and the protein-monomer complexes, or via including new parameters that multiply certain terms in the equations of motion \cite{bindschadler2004mechanistic, dufort1996profilin, yarmola2008effect}.  These adaptations could be straightforwardly included in the modeling done here.  Additionally, the effect of different concentrations of solvated ATP, ADP-Pi, and ADP could be investigated by changing the values of the pseudo-first order reaction rate constants, or even by regarding those reactions as second order and tracking the concentrations of the nucleotide species as separate variables.  These modifications run counter to the goal here of model reduction, however.  Different choices in modeling are of course suited to different purposes, and the choices made here address a desire to have a simple mental picture of an otherwise obscured and complex process.

\appendix
\section*{Appendix A: Details of BC Model}
The BC model consists of the following set of 11 coupled ordinary differential equations:
\begin{eqnarray}
\label{eq29}
&&\frac{d G^T}{dt} = k^T_\text{off, +} T^T_+ + k^T_\text{off, - } T^T_- + k_\text{nex} G^D - k_\text{renex} G^T - G^T N (k^T_\text{on, +} + k^T_\text{on, -})  \\
\label{eq30}
&& \frac{d G^D}{dt} =  k^D_\text{off, +} T^D_+ + k^D_\text{off, - } T^D_- + k^{Pi}_\text{off, +}T^{Pi}_+ + k^{Pi}_\text{off, -} T^{Pi}_- + k_\text{renex} G^T - k_\text{nex} G^D- \\
&& \ \ \ \ \ \ \ \ \ G^D N (k^D_\text{on, +} + k^D_\text{on, -}) \notag  \\
\label{eq31}
&&\frac{d F^T}{dt} = G^T N (k^T_\text{on, +} + k^T_\text{on, -}) - k^T_\text{off, +} T^T_+ - k^T_\text{off, - }T^T_- - k_\text{hyd} F^T + k_\text{rehyd} F^{Pi}\\
\label{eq32}
&&\frac{d F^{Pi}}{dt} = k_\text{hyd}F^T - k_\text{rehyd}F^{Pi} - k^{Pi}_\text{off, +} T^{Pi}_+ - k^{Pi}_\text{off, -}T^{Pi}_- - k_\text{phos}F^{Pi} + k_\text{rephos} F^D\\
\label{eq33}
&&\frac{d F^D}{dt} = G^D N (k^D_\text{on, +} + k^D_\text{on, -})+ k_\text{phos}F^{Pi} - k_\text{rephos}F^D-k^D_\text{off, +}T^D_+ - k^D_\text{off, -}T^D_-\\
\label{eq34}
&&\frac{d T^T_\pm}{dt} = k^T_\text{on, $\pm$} G^T (T^{Pi}_\pm + T^D_\pm) +(k^{Pi}_\text{off, $\pm$} T^{Pi}_\pm + k^D_\text{off, $\pm$}T^D_\pm)\eta^T_\pm -k_\text{hyd}T^T_\pm + k_\text{rehyd}T^{Pi}_\pm \\
&& \ \ \ \ \ \ \ \ - k^T_\text{off, $\pm$}T^T_\pm(\eta^{Pi}_\pm + \eta^D_\pm) - k^D_\text{on, $\pm$} G^D T^T_\pm \notag  \\
\label{eq35}
&&\frac{d T^{Pi}_\pm}{dt} = k_\text{hyd}T^T_\pm - k_\text{rehyd}T^{Pi}_\pm + (k^T_\text{off, $\pm$} T^T_\pm + k^D_\text{off, $\pm$} T^D_\pm) \eta^{Pi}_\pm - k_\text{phos}T^{Pi}_\pm +k_\text{rephos} T^D_\pm \\
&& \ \ \ \ \ \ \ \ - k^{Pi}_\text{off, $\pm$} T^{Pi}_\pm (\eta^T_\pm + \eta^D_\pm) - T^{Pi}_\pm(k^T_\text{on, $\pm$} G^T + k^D_\text{on, $\pm$} G^D) \notag  \\
\label{eq36}
&&\frac{d T^D_\pm}{dt} =  k^D_\text{on, $\pm$} G^D ( T^T_\pm + T^{Pi}_\pm)  + (k^T_\text{off, $\pm$} T^T_\pm + k^{Pi}_\text{off, $\pm$} T^{Pi}_\pm) \eta^D_\pm + k_\text{phos}T^{Pi}_\pm - k_\text{rephos}T^D_\pm\\
&& \ \ \ \ \ \ \ \ \  - k^D_\text{off, $\pm$}T^D_\pm(\eta^T_\pm + \eta^{Pi}_\pm) - k^T_\text{on, $\pm$} G^T T^D_\pm \notag 
\end{eqnarray}
where
\begin{eqnarray}
\label{eq37}
&&\eta^T_\pm = \frac{T^T_\pm}{N}\bigg(1 - \frac{T^{Pi}_\pm}{N}\bigg)\\
\label{eq38}
&&\eta^D_\pm = \frac{T^D_\pm}{N}\\
\label{e39}
&&\eta^{Pi}_\pm = 1 - \eta^T_\pm - \eta^D_\pm
\end{eqnarray}
In Table 1 we list the meaning and values of the rate constants used in our implementation of the BC model.  
\begin{center}
\begin{table}[H]
\centering
\begin{tabular}{|c|c|c|}
\hline
Label & Reaction & Value\\
\Xhline{3\arrayrulewidth}
$k^T_\text{on, +}$ & Polymerization of $G^T$ to barbed end & 11.6 $\mu \text{M}^{-1} \ \text{s}^{-1}$\\
\hline
$k^T_\text{on, -}$ & Polymerization of $G^T$ to pointed end & 1.3 $\mu \text{M}^{-1} \ \text{s}^{-1}$\\
\hline
$k^T_\text{off, +}$ & Depolymerization of $F^T$ from barbed end & 1.4 $ \text{s}^{-1}$\\
\hline
$k^T_\text{off, -}$ & Depolymerization of $F^T$ from pointed end  & 0.8 $ \text{s}^{-1}$\\
\hline
$k^D_\text{on, +}$ & Polymerization of $G^D$ to barbed end & 2.9 $\mu \text{M}^{-1} \ \text{s}^{-1}$\\
\hline
$k^D_\text{on, -}$ & Polymerization of $G^D$ to pointed end & 0.13 $\mu \text{M}^{-1} \ \text{s}^{-1}$\\
\hline
$k^D_\text{off, +}$ & Depolymerization of $F^D$ from barbed end & 5.4$ \ \text{s}^{-1}$\\
\hline
$k^D_\text{off, -}$ & Depolymerization of $F^D$ from pointed end & 0.25$ \  \text{s}^{-1}$\\
\hline
$k^{Pi}_\text{off, +}$ & Depolymerization of $F^{Pi}$ from barbed end & 1.4$ \  \text{s}^{-1}$\\
\hline
$k^{Pi}_\text{off, -}$ & Depolymerization of $F^{Pi}$ from pointed end & 0.8$ \  \text{s}^{-1}$\\
\hline
$k^\text{hyd} $ & ATP hydrolysis converting $F^T$ to $F^{Pi}$ & 0.3$ \  \text{s}^{-1}$\\
\hline
$k^\text{rehyd} $&  ATP condensation converting $F^{Pi}$ to $F^T$ & $  \epsilon$ 0.3$ \ \text{s}^{-1}$\\
\hline
$k^\text{nex} $ & Nucleotide exchange converting $G^D$ to $G^T$ & 0.01 $\text{s}^{-1}$\\
\hline
$k^\text{renex} $ & Nucleotide exchange converting $G^T$ to $G^D$ & $\epsilon$ 0.01 $\text{s}^{-1}$\\
\hline
$k^\text{phos} $ & Inorganic phosphate release converting $F^{Pi}$ to $F^D$ & 0.002 $\text{s}^{-1}$\\
\hline
$k^\text{phos} $ & Inorganic phosphate capture converting $F^D$ to $F^{Pi}$ &$\epsilon$ 0.002 $\text{s}^{-1}$\\
\hline
\end{tabular}
\caption{Rate constants in the BC model.  The prefix ``re" indicates the reverse of a kinetically dominant forward reaction (i.e. nearly irreversible reactions).  The value of rate constants for these reverse reactions is taken to be equal to the corresponding forward reaction rate multiplied by a small parameter $\epsilon$.  We typically take $\epsilon = 0.01$.  Some of these reactions, such as the nucleotide exchange reaction, are second order reactions.  For example the proper rate of reaction for conversion of $G^D$ to $G^T$ is $k^*_\text{nex}[G^D][ATP]$.  We treat such cases as pseudo-first order reactions by assuming that the concentration of the species which we don't track is constant and that its concentration is contained in the rate constant used.  Thus $k_\text{nex} = k^*_\text{nex}[ATP]$ in our model.  This assumption of constant concentrations of free ATP, ADP, and Pi is reasonable in cellular environments.   All values are taken from \cite{brooks2009nonequilibrium}.}
\end{table}

\end{center}

We note that the original equations in the BC model did not include reversible reactions as shown here.  This amounts to setting $k_\text{renex}$, $k_\text{rehyd}$, and $k_\text{rephos}$ to 0 Equations \ref{eq29}-\ref{eq36}.  The interpretation of $\eta^i_j$ is the probability that the monomer adjacent to the $j$ tip is in the $i$ hydrolysis state.  The equations of motion of these variables in principle depend on the hydrolysis state of the monomer next to them toward the center of the filament, and Equations \ref{eq29}-\ref{eq36} represent the truncation of this set of recursion equations.  This is accomplished by assuming that $\eta^i_j$ depends only on the tip monomer hydrolysis state through Equations \ref{eq34}-\ref{eq36}.  These equations were arrived at by inspecting the time course of the $\eta^i_j$ and the $T^i_j$ variables and discerning the equations which approximately related the two.  The system of Equations \ref{eq29}-\ref{eq36} does not admit and analytical solution but can be numerically integrated with appropriate initial conditions specified \cite{brooks2009nonequilibrium}.    

\begin{figure}[H]
\centering
\includegraphics[width=12 cm]{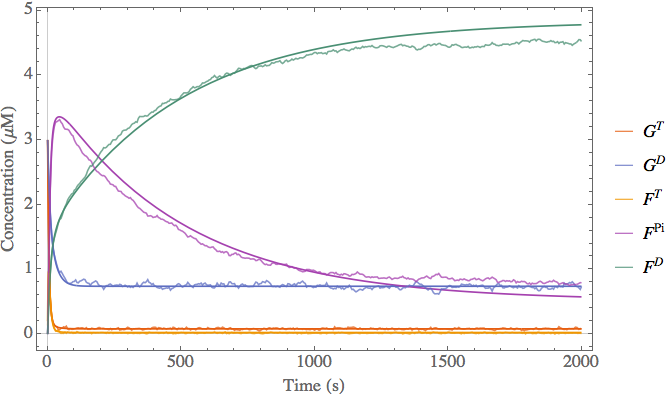}
\caption{Time course of the concentrations of the various species following the addition of 3 $\mu \text{M}$ of $G^T$ and 3 $\mu \text{M}$ of $G^D$  actin to a bath with a number concentration $N=0.017$ $\mu \text{M}$ of seed filaments.  The mean-field BC model accurately predicts the shape of the time-courses resulting from the stochastic simulation.}
\label{fig4}
\end{figure}

\begin{figure}[H]
\centering
\includegraphics[width=13 cm]{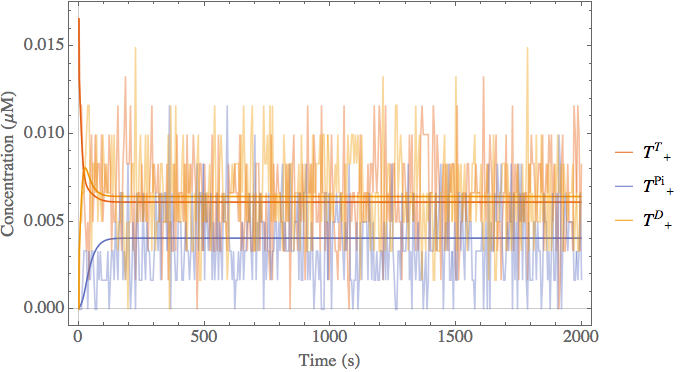}
\caption{Time course of the concentrations of the various tip species following the addition of 3 $\mu \text{M}$ of $G^T$ and 3 $\mu \text{M}$ of $G^D$  actin to a bath with a number concentration $N=0.017$ $\mu \text{M}$ of seed filaments.  The noisiness of the stochastic trajectories results from the small copy number of filaments.}
\label{fig5a}
\end{figure}

\begin{figure}[H]
\centering
\includegraphics[width=13 cm]{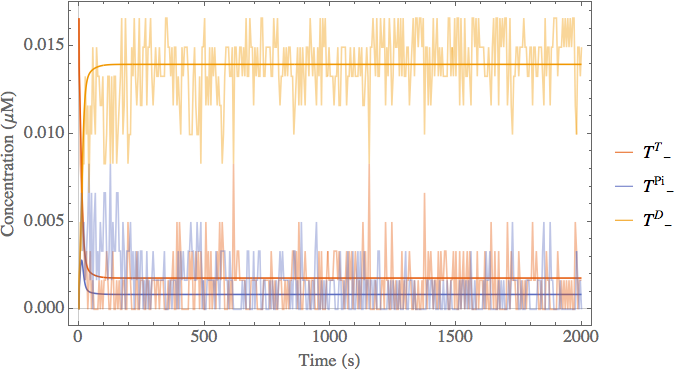}
\caption{Time course of the concentrations of the various tip species following the addition of 3 $\mu \text{M}$ of $G^T$ and 3 $\mu \text{M}$ of $G^D$  actin to a bath with a number concentration $N=0.017$ $\mu \text{M}$ of seed filaments.  The noisiness of the stochastic trajectories results from the small copy number of filaments.}
\label{fig5b}
\end{figure}

We check the accuracy of the truncation assumption of the BC model by comparing a predicted time course to the results of a simulation using the software package MEDYAN (Mechanochemical Dynamics of Active Networks).  MEDYAN was developed to perform coarse-grained simulations of active networks and combines stochastic chemical algorithms with detailed mechanics as well as coupling between reaction rates of force-sensitive chemical reactions and the mechanical state of the species involved \cite{popov2016medyan}.  In Figures \ref{fig4}, \ref{fig5a}, and \ref{fig5b} we show the simulated time courses as well as the mean-field prediction of the BC model.

The steady-state vector $\bold{x}^\text{ss}$ satisfying $\bold{f}(\bold{x}^\text{ss}) = \bold{0}$ can be found by numerically finding the root of the right and sides of Equations \ref{eq29}-\ref{eq36}.   Equations \ref{eq29}-\ref{eq33} sum to zero, reflecting the conservation of total actin $M$ encoded in these reactions. Also each of the two sets of Equations \ref{eq34}-\ref{eq36} sum to zero, reflecting the conservation of plus end tips and minus end tips.  Thus the system $\dot{\bold{x}} = \bold{f}(\bold{x})$ represented by Equations \ref{eq29}-\ref{eq36} is linearly dependent, and no solution to exists $\bold{f}(\bold{x}^\text{ss}) = 0$ unless we specify additional equations.  These additional equations are 
\begin{eqnarray}
\label{eq40}
&& G^T + G^D + F^T + F^{Pi} + F^D = M\\
\label{eq41}
&& T^T_\pm + T^{Pi}_\pm + T^D_\pm = N_\pm
\end{eqnarray}
For linear filaments considered here, the number of plus end tips is equal to the number of minus end tips: $N_+ = N_- = N$.  Solving $\bold{f}(\bold{x}^\text{ss}) = 0$ gives one solution for which all variables are nonnegative, so there is a unique realistic steady-state solution for a given set of parameters $M$ and $N$.  The eigenvalues of the BC Jacobian evaluated at the equilibrium point $\bold{J^*} = \frac{\partial \bold{f}}{\partial \bold{x}}\big\vert _{\bold{x} = \bold{x}^\text{ss}}$ indicate the stability of the steady-state.  3 of the 11 eigenvalues are zero, corresponding to the 3 linear conservation laws in our system.  This implies that the dynamics of the BC model lie on an 8-dimensional submanifold of the full 11-dimensional variable space, and this submanifold is determined by the parameters $M$ and $N$.  The rest of the eigenvalues are negative, indicating that the unique nonnegative vector $\bold{x}^\text{ss}$ is attracting and stable.  We note that this equilibrium point of the dynamics actually corresponds to a non-equilibrium steady-state of the chemical system, since this state corresponds to actin treadmilling in which there are equal and opposite rates of polymerization at the barbed and pointed ends of the filaments, fueled by ATP hydrolysis.  

\section*{Appendix B: Steady-State Concentrations in the BC Model}
\begin{figure}[H]
\centering
\includegraphics[width=10 cm]{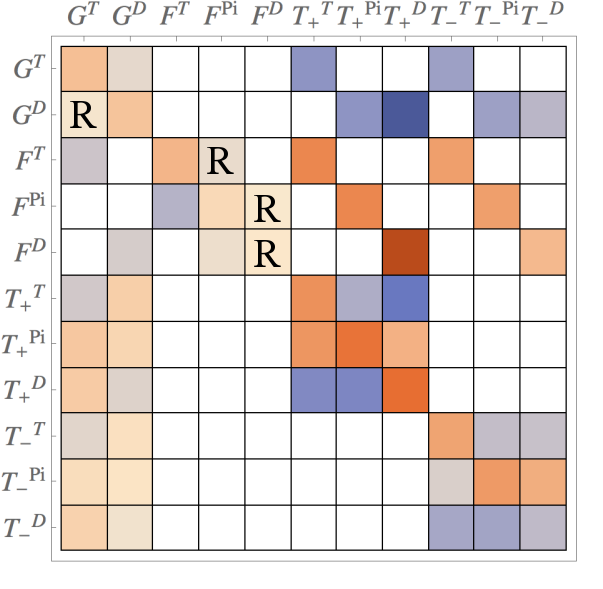}
\caption{Visualization of $\bold{J^*}$ to understand the couplings in the BC model.  If the evolution of species $\bold{x}_i$ depends on species $\bold{x}_j$, then $J^*_{ij}$ will be nonzero.  If this coupling causes $\bold{x}_i$ to increase, the entry in the matrix is colored blue, and if it causes $\bold{x}_i$ to decrease, the entry is red.  The saturation of the color indicates the magnitude of the coupling constant.  An ``R" is placed in the plot if that element is nonzero only due to the inclusion of slow reversible reactions, when $\epsilon >0$.  The corresponding plot for the CT model is identical to the top left 5 by 5 matrix in this plot.  }
\label{fig11}
\end{figure}

We show here that the feature of the lack of dependence of the steady-state concentrations of some species on $M$ is also present in the BC model.  We do this by consideration of the system of algebraic equations $\bold{f}(\bold{x}^\text{ss})=\bold{0}$.  The Jacobian of system at the steady-state is effectively visualized in Figure \ref{fig11}.   
Now that we can grasp which species are coupled to which other species, we observe that no species depend on $F^T$, $F^{Pi}$, or $F^D$, except those species themselves.  Therefore we could ignore those equations and the resulting subsystem of equations would be closed.  Now that subsystem will be linearly dependent, but we can supplement it with the following equations to make it independent:
\begin{eqnarray}
&& T^T_\pm + T^{Pi}_\pm + T^D_\pm = N
\end{eqnarray} 
The subsystem of equations could now be solved, giving the steady-state concentrations of all species except for $F^T$, $F^{Pi}$, and $F^D$.  $M$, the total amount of actin, does not enter into any of the equations of the subsystem, and so the solution of that system does not depend on $M$.  Therefore, the steady-state concentration of each species except $F^T$, $F^{Pi}$, and $F^D$ does not depend on the initial conditions, however they do depend on the parameter $N$.  

Now we consider the case where $\epsilon = 0$.  Referring to Figure \ref{fig11}, we see that this would mean that no species depends on $F^D$.  By reasoning similar to the above, this implies that the steady-state concentration of each species except $F^D$ could be determined without specifying $M$.  Thus, the steady-state concentration of only $F^D$ depends on the initial conditions in this case.  All the arguments here apply to the CT model as well.   

\section*{Acknowledgements}
We would like to thank Maria Cameron, for her helpful comments regarding the diffusion map analysis presented in this work, and Andrew Maven Smith, for his insights into the solution of the CT model, among other things.  This work is partially supported by the National Science Foundation NSF CHE-136081.

\bibliographystyle{unsrt}

\end{document}


\begin{center}
\begin{huge} 
\textbf{Supporting Information for: Low-Dimensional Dynamics of Actin Polymerization}
\end{huge}

\begin{large}
\text{Carlos Floyd, Christopher Jarzynski, Garegin Papoian}
\end{large}
\end{center}
\title{Supporting Information for: Low-Dimensional Dynamics of Actin Polymerization}
\author{Carlos Floyd, Christopher Jarzynski, Garegin Papoian}
\date{August 2017}

\section*{Supporting Information 1: Diffusion Map Analysis of the BC Model}
The analysis presented here is based primarily off the method described in \cite{singer2009detecting}, in which the application of anisotropic diffusion maps allows one to find intrinsic low-dimensional manifolds of high-dimensional data associated with a dynamical system.  In general, diffusion mapping is a nonlinear technique to rearrange data based on some intrinsic features of the underlying geometry.  In order to do this analysis, one needs a data set, not a numerical solution to an ODE system, and such a data set for this system can be obtained by a numerical integration of a Chemical Langevin Equation (CLE), a stochastic differential equation describing the time evolution of a chemical system \cite{gillespie2000chemical}.  A data set generated this way has the same average properties of the corresponding deterministic system as well as fluctuations corresponding to the stochastic nature of reaction occurrence.  

For a system of $n$ chemical species reacting through $m$ reaction channels, we define the time-dependent $n\times 1$ vector of number of species $ \bold{X}(t)$, the $m\times 1$ reaction propensity vector function, $\bold{a}(\bold{X}(t))$, which determines the rate of the $m$ reactions as a function of the concentrations of the $n$ species, and the $n \times m$ state-change matrix $\bold{v}$, which encodes the stoichiometry of the $m$ reactions.  The quantity $a_i(\bold{X}(t))dt$ is the probability that one instance reaction $i$ will occur in the time interval $[t,t+dt)$ given the vector of species number $\bold{X}(t)$.  The quantity $v_{i,j}$ is the change in the number of species $i$ produced during reaction $j$.  The CLE can be written as
\begin{equation}
\label{eq41}
\frac{d X_i(t)}{dt} = \sum_{j=1}^m v_{i,j} a_j(\bold{X}(t)) + \sum_{j=1}^m v_{i,j} a_j^{1/2}(\bold{X}(t)) \gamma_j(t)
\end{equation}
where $\gamma_j(t)$ are independent and temporally uncorrelated Gaussian white noises.  More compactly, the CLE can be written in the It\^{o} form as 
\begin{equation}
\label{eq42}
d\bold{X}(t) = \bold{v}\bold{a}dt + \bold{v} (\bold{a}^{\circ 1/2} \circ d \bold{w})
\end{equation}
where $\circ$ denotes the Hadamard, or element-wise, matrix multiplication or matrix power, $\bold{w}$ is a $ m\times 1$ vector of standard independent Brownian motions, and we have dropped the explicit dependence of $\bold{a}$ on $\bold{X}(t)$.  An Euler-Maruyama integration of the CLE generates a stochastic data set of the trajectory of the chemical system.

With a data set in hand, one can perform anisotropic diffusion mapping on it.  Broadly, the goal of such mapping is to determine for each pair of points in the data set the distance between them, where the distance is defined to account for the anisotropic likelihood of moving in certain directions in a certain time step.  Then one defines a Markov jump process between these points with jump probabilities determined by the anisotropic distances.  Eigendecomposition of the jump matrix will indicate the number of dimensions that are necessary to capture the jump dynamics with sufficient accuracy, hopefully with result that dimensionality reduction is achieved.  The dominant eigenvectors of the jump matrix form a basis for the diffusion space; the $i^{th}$ data point gets mapped to a vector containing containing the $i^{th}$ coordinate of each of the dominant eigenvectors.  Thus distance between two data points in the diffusion space corresponds to the probability to jump from one data point to the other in the Markov jump process.  We refer the reader to \cite{bah2008diffusion, sipola2013knowledge, singer2009detecting} for more details.  

For the BC model, $n = 11$ and $m=50$.  We generate a data set of 2,000 points by running a trajectory initialized at the species number vector $(1.8 \times 10 ^5, 7.2 \times 10^5, 9.0 \times 10^4, 9.0 \times 10^4, 1.8 \times 10^5, 6.0 \times 10^3, 6.0 \times 10^3, 6.0 \times 10^3, 6.0 \times 10^3, 6.0 \times 10^3, 6.0 \times 10^3)^\intercal$ for 0.02 s with integration time steps of $h=1 \times 10^{-5}$ s.  For each data point $\bold{x}^{(i)}$, a collection ($10^3$) of short simulation bursts is generated and the covariance matrix $\bold{\Sigma}^{(i)}$ is calculated based on these statistics.  These enter into the calculation of the anisotropic distance $d^2_\bold{\Sigma}(\bold{x}^{(i)},\bold{x}^{(j)})$, a $2,000 \times 2,000$ matrix.  The weight matrix $\bold{W}$ based on these distances is calculated as 

\begin{equation}
\label{eq43}
W_{i,j} = e^{-d^2_\bold{\Sigma}(\bold{x}^{(i)},\bold{x}^{(j)})/ \epsilon}
\end{equation}
where

\begin{equation}
\label{eq43a}
d^2_\bold{\Sigma}(\bold{x}^{(i)},\bold{x}^{(j)}) = \frac{1}{2}(\bold{x}^{(i)}-\bold{x}^{(j)})^\intercal \big((\Sigma^{(i)})^{-1}+(\Sigma^{(j)})^{-1}\big)(\bold{x}^{(i)}-\bold{x}^{(j)})
\end{equation}
(Note: here we define $\bold{W}$ using $\epsilon$ in the denominator of the exponent, but elsewhere in the literature the denominator might appear as $\epsilon^2$.)  Multiple such data sets for a given set of initial conditions are generated to confirm the robustness of the results.  The choice of parameter $\epsilon$ is not trivial and is discussed below.  $\bold{W}$ is normalized by its row sums to convert it to a row stochastic matrix $\bold{\mathcal{A}}$ via
\begin{equation}
\label{eq44}
\bold{\mathcal{A}} = \bold{D}^{-1}\bold{W}
\end{equation}
where $\bold{D}$ is a diagonal matrix whose $i,i^{th}$ element is equal to the sum of the $i^{th}$ row of $\bold{W}$.  The symmetric matrix $\bold{S} = \bold{D}^{-1/2}\bold{W}\bold{D}^{-1/2}$ is similar to $\bold{\mathcal{A}}$, since $\bold{\mathcal{A}} = \bold{D}^{-1/2}\bold{S}\bold{D}^{1/2}$.   Eigendecomposition on $\bold{S}$ is done to find the dominant eigenvectors which will form the basis of the diffusion space.  

\begin{figure}[H]
\centering
\includegraphics[width=13 cm]{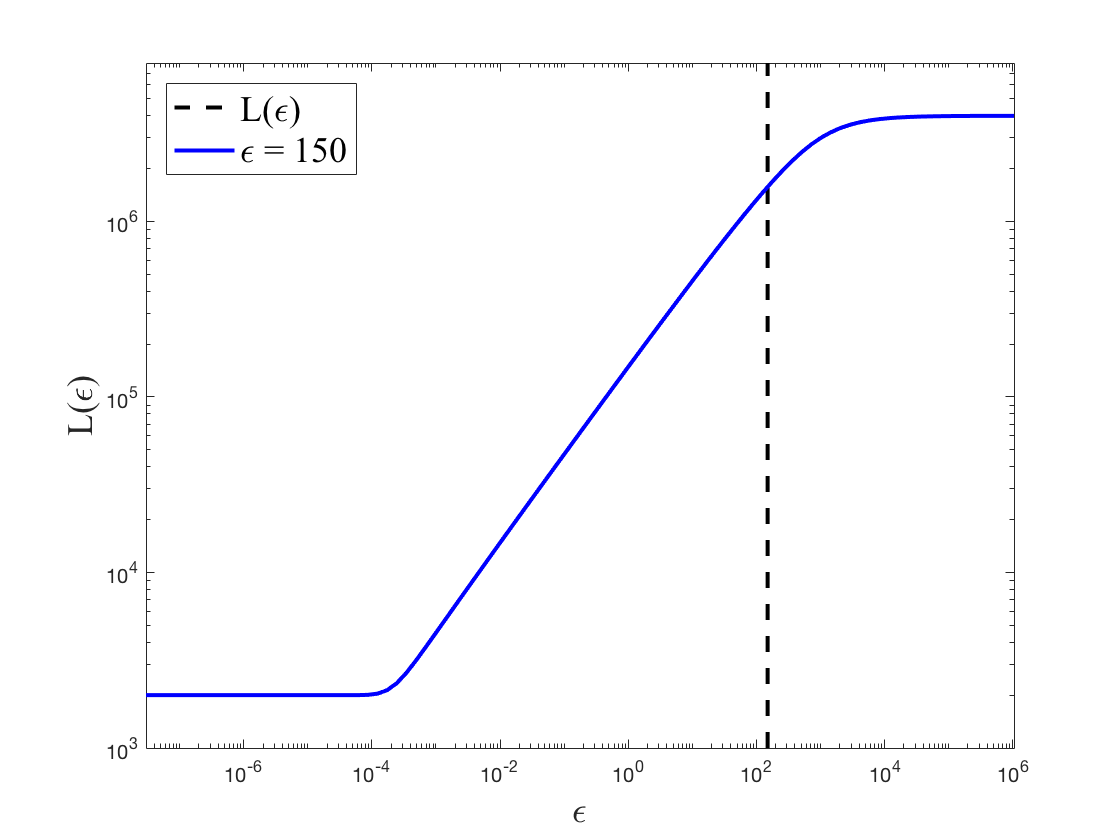}
\caption{The function $L(\epsilon)$ appears linear on a log-log scale for a range of values of $\epsilon$, including the choice of $\epsilon$ as the median of the distance matrix $d^2_\bold{\Sigma}$.  This choice of $\epsilon$ seems to lie at the edge of the domain where $L(\epsilon)$ appears linear, however we found that the diffusion map results obtained are robust against smaller values of $\epsilon$ being chosen.  Qualitatively, the results are the same for choices of $\epsilon$ as low as 10 (data not shown).}
\label{fig6}
\end{figure}

Choosing $\epsilon$ amounts to setting the scale of the diffusion process.  Several heuristics exist for determining this parameter given ones data set \cite{sipola2013knowledge}.  One such heuristic is to choose it as the median of the distance matrix:
\begin{equation}
\label{eq45}
\epsilon = \text{median}\{d^2_\bold{\Sigma}\}
\end{equation}
For our data set, $\text{median}\{d^2_\bold{\Sigma}\} \approx 150$.  
Another heuristic is to plot the function $L(\epsilon) = \sum_i \sum_j W_{i,j}$ on a log-log scale and choose $\epsilon$ from the region where this graph is linear.  For our data set, these heuristics are compatible, as illustrated in Figure \ref{fig6}.

\begin{figure}[H]
\centering
\includegraphics[width=13 cm]{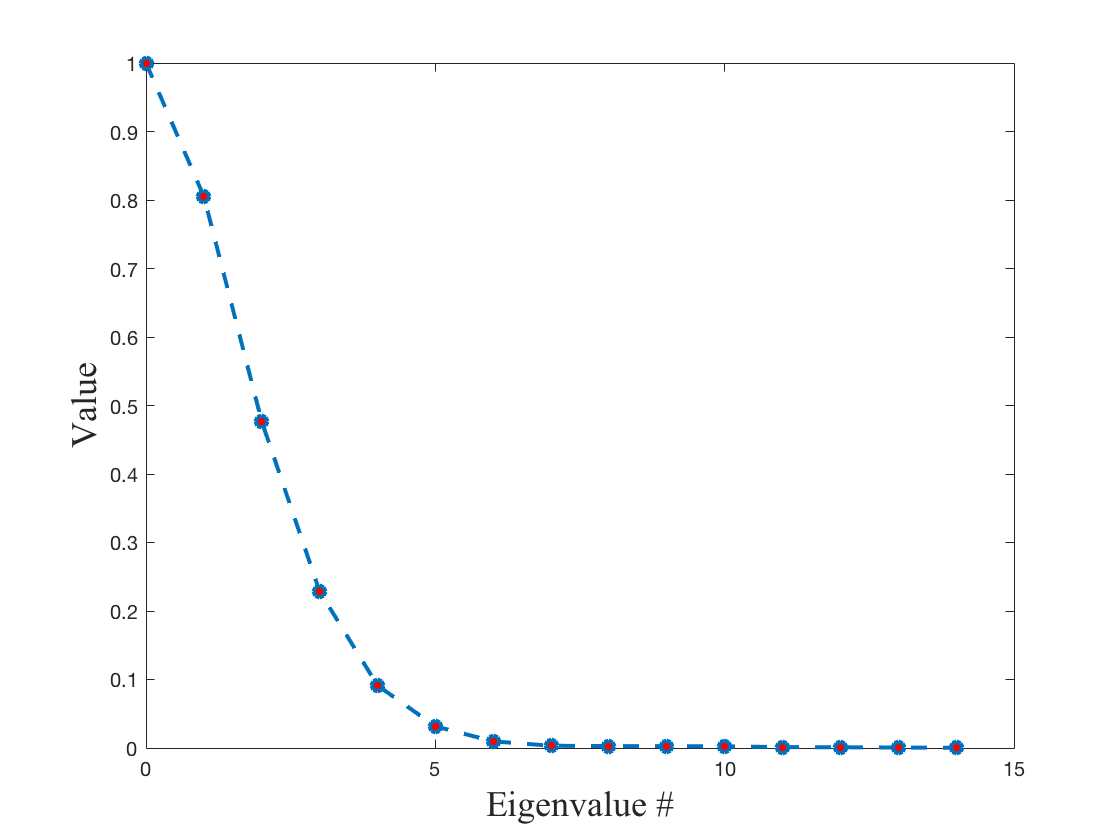}
\caption{The eigenspectrum of $\bold{S}$ indicates that only a couple of dimensions are important in describing the diffusion process corresponding to the stochastic matrix $\bold{S}$, and by extension the dynamics of the BC model.}
\label{fig7}
\end{figure}

The eigenspectrum of $\bold{S}$ is shown in Figure \ref{fig7}.  The first eigenvalue $\lambda_0 =1$ is trivial, and the successive eigenvalues are all less than 1.  The number of eigenvalues with appreciable magnitudes indicate the number of dimensions needed to describe diffusion according to the stochastic matrix $\bold{S}$ with sufficient accuracy.  

\begin{figure}[H]
\centering
\includegraphics[width=13 cm]{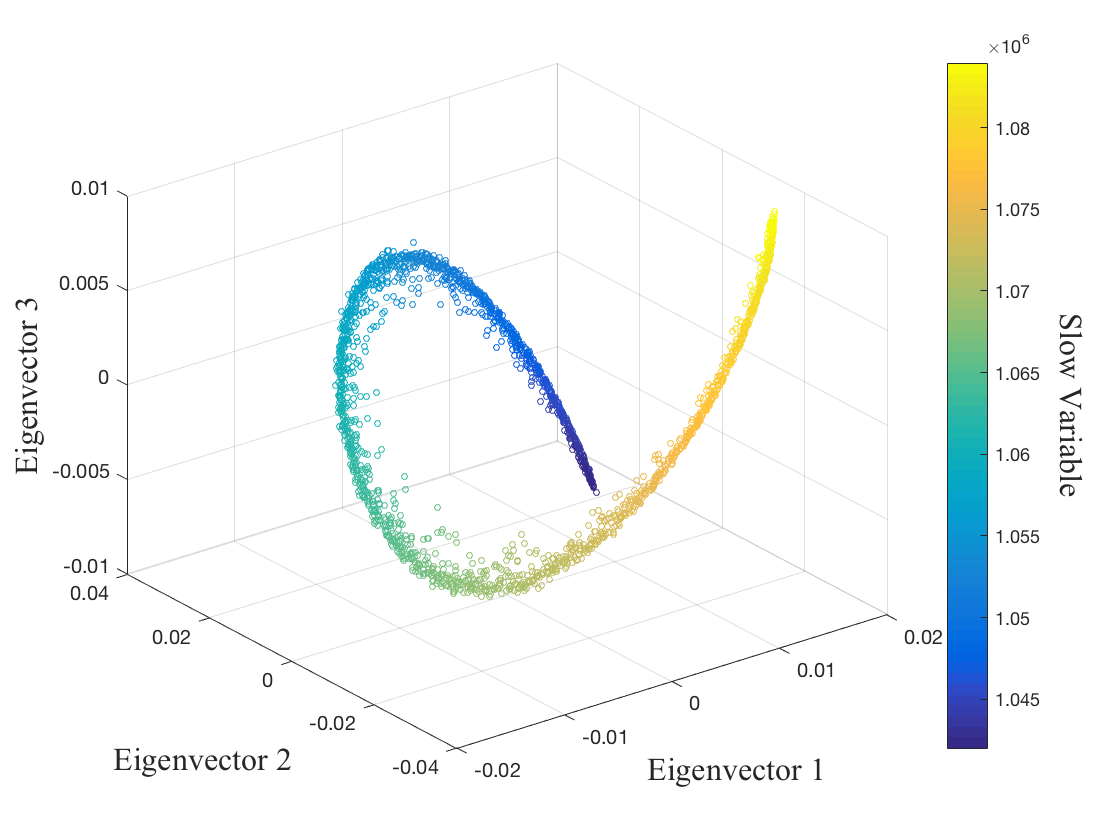}
\caption{The diffusion mapped data set lies more or less on a smooth non-intersecting curve, indicating that the mapping is continuous and one-to-one.  Furthermore, the slow variable $G^T(t) + G^D(t) + F^T(t) +F^{Pi}(t)$ is a relatively uniform and monotonic parameter for the curve.  This indicates that the mapping has captured the slow, low-dimensional dynamics of the data set.  }
\label{fig8}
\end{figure}

Apparently only a small number of dimensions are necessary to describe the dynamics in the diffusion space.  We illustrate this by choosing the top 3 eigenvectors as the basis of the diffusion space.  The mapped data in this space should lie on a smooth, non-intersecting curve.  Furthermore, the curve should be able to describe the dynamics of a ``slow variable," that is, a collective variable whose evolution take place on the slowest time-scale.  In Figure \ref{fig8}, we display the diffusion mapping, as parameterized by the slow variable $G^T(t) + G^D(t) + F^T(t) +F^{Pi}(t)$.  

\begin{figure}[H]
\centering
\includegraphics[width=13 cm]{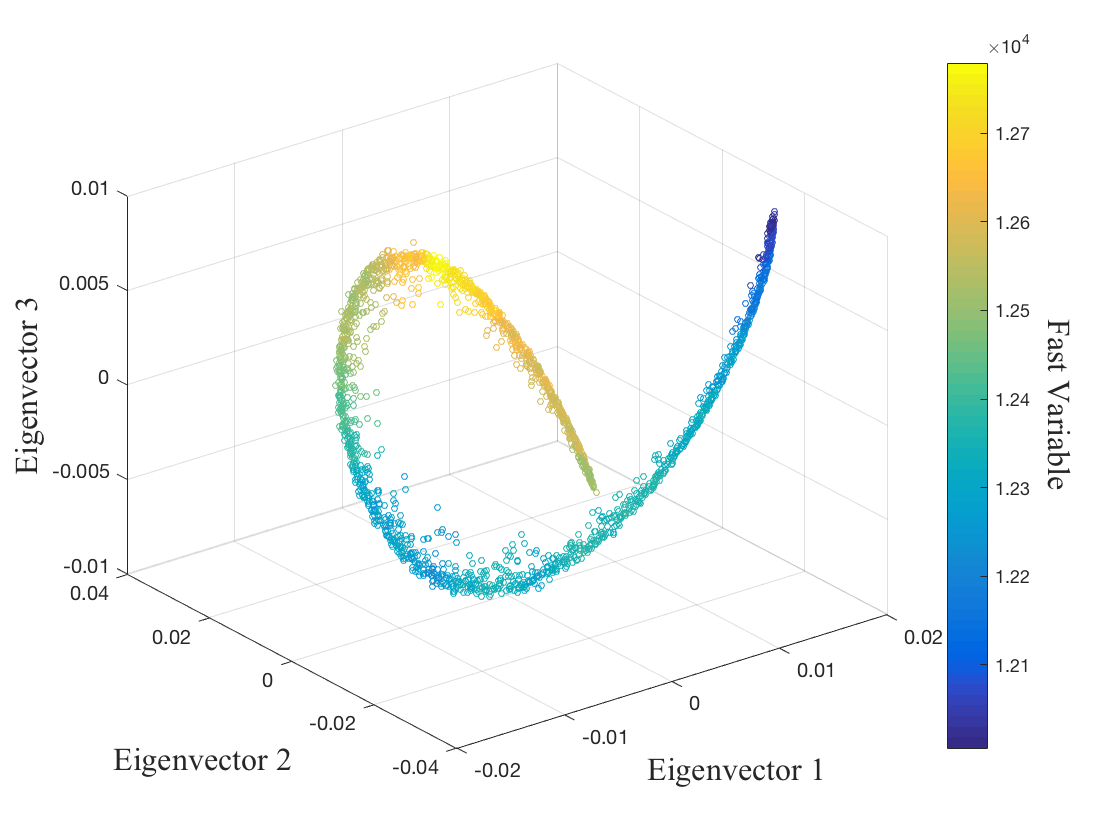}
\caption{The parameterization of the data set by the fast variable $T_-^T(t) + T_-^{Pi}(t)$ is non-monotonic and uneven, indicating that this fast process is not captured by the low-dimensional description of the diffusion mapping.}
\label{fig9}
\end{figure}

For comparison, we show in Figure \ref{fig9}  a diffusion mapping parameterized by the fast variable $T_-^T(t) + T_-^{Pi}(t)$. 

\begin{figure}[H]
\centering
\includegraphics[width=13 cm]{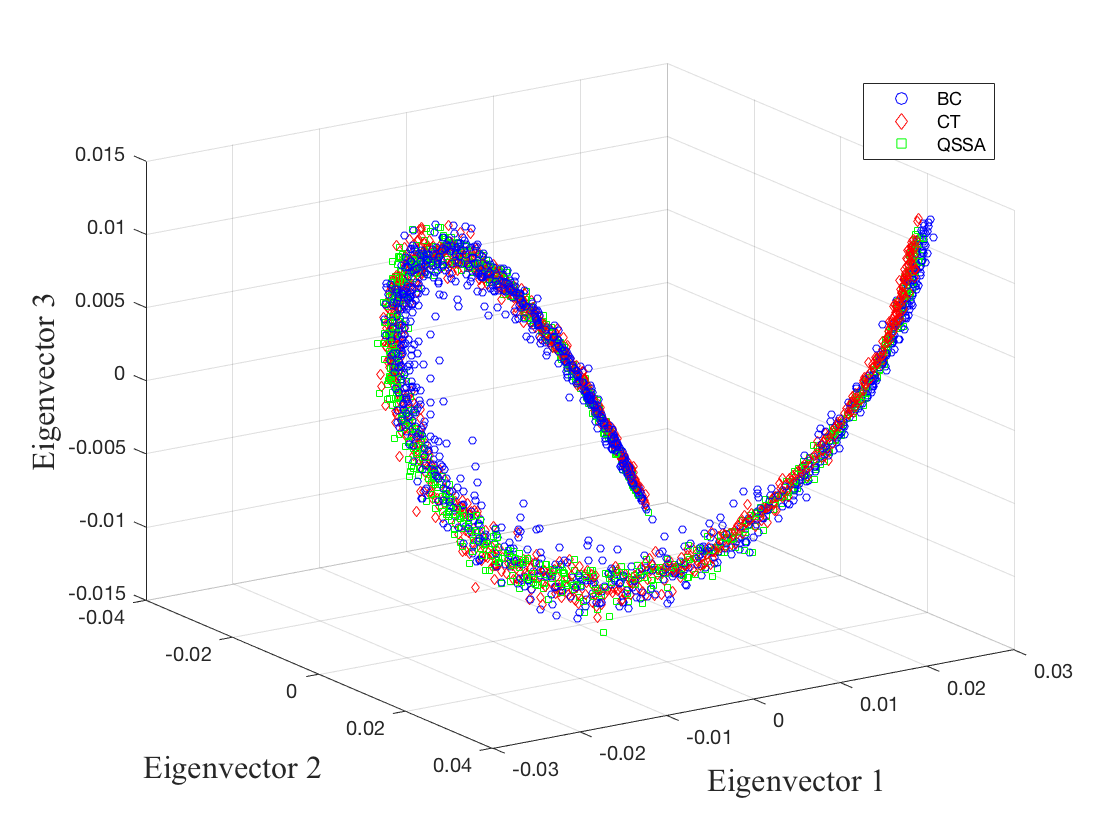}
\caption{The diffusion mapped data sets using the BC, QSSA, and CT models all lie on the same curve.  The implication is that all of these models represent the same low-dimensional dynamics of interest.}
\label{fig10}
\end{figure}

The fact that the smooth curve in Figure \ref{fig8} is parameterized well by the dynamics of the slow collective variable indicates that the diffusion mapping has nicely captured the slow, low-dimensional dynamics of the BC model.  Only three dimensions sufficed for this description, indicating that the BC model dynamics can be approximately described by a lower-dimensional system, and model reduction to this effect is justified.  Lastly, if we repeat this analysis for the same initial conditions using the QSSA model and CT model, we find that the curves in the diffusion space agree, as indicated in Figure \ref{fig10}.  

\section*{Supporting Information 2: Solution of Equation 10 Using the Drazin Inverse}
As mentioned in the main text, the solution of 
\begin{equation}
\label{eq47}
\dot{\bold{x}} = \bold{A}\bold{x}+\bold{b}
\end{equation}
is somewhat complicated by the fact that $\bold{A}$ is singular, owing to the presence of linear conservation laws, which are common in chemical reaction networks.  Options for solving Equation \ref{eq47} include eliminating variables until the dimension of $\bold{A}$ is equal to its rank, however this approach requires one to rewrite $\bold{A}$ in such a way that the interpretation of its entries as rates of certain reactions is obscured.  Also, one could do a linear transformation of $\bold{A}$ to diagonalize it, but then one has a solution in terms of linear combinations of the variables, and the meaning of these linear combinations is not always intuitive.  We propose a method involving the Drazin inverse $\bold{A}^\mathcal{D}$ as an alternative that avoids these potential problems.  We note that the machinery of the Drazin inverse is slight overkill for solving the CT model, and in fact using the Penrose inverse would allow one to solve the system cleanly as well.  The Drazin inverse method is especially useful if $\bold{A}$ would be non-diagonalizable in addition to singular.  We illustrate this method here for the purpose of its broader usefulness in other applications.  Using Equations \ref{eq55}, \ref{eq57}, and \ref{eq58}, one can write the solution of Equation \ref{eq48} immediately, and the result is in terms of the original variables. 

The general solution of Equation \ref{eq47} is
\begin{equation}
\label{eq48}
\bold{x} = e^{\bold{A}t}\bigg(\int e^{-\bold{A}t}\bold{b} \ \text{dt}\bigg).
\end{equation}
If $\bold{A}$ is nonsingular and $\bold{b}$ is constant, then 
\begin{equation}
\label{eq49}
\int e^{-\bold{A}t}\bold{b} \ \text{dt} = -\bold{A}^{-1}(e^{-\bold{A}t} - \bold{I})\bold{b}
\end{equation}
We now introduce the Drazin inverse $\bold{A}^\mathcal{D}$ before we discuss its use in solving Equation \ref{eq47} for singular $\bold{A}$.  The Drazin inverse is defined as the unique solution of 
the following equations
\begin{eqnarray}
\label{eq50}
&\bold{A}\bold{A}^\mathcal{D}=\bold{A}^\mathcal{D}\bold{A} \\
\label{eq51}
& \bold{A}^\mathcal{D}\bold{A}\bold{A}^\mathcal{D} = \bold{A}^\mathcal{D}\\
\label{eq52}
& \bold{A}^\mathcal{D}\bold{A}^{k+1} = \bold{A}^k
\end{eqnarray}
Here $k$ is the index of the matrix, defined as the smallest non-negative integer for which 
\begin{equation}
\label{eq53}
\text{rank}(\bold{A}^{k+1})=\text{rank}(\bold{A}^k)
\end{equation}
Note that for a nonsingular matrix, $k = 0$, and $\bold{A}^\mathcal{D} = \bold{A}^{-1}$ is a solution of Equations \ref{eq50}-\ref{eq52}.  In general, if $\bold{A}$ has Jordan decomposition 
\begin{equation*}
\bold{A} = \bold{V}
\begin{pmatrix}
J_1 & 0\\
0 & J_0\\
\end{pmatrix}
\bold{V}^{-1} 
\end{equation*}
where $J_1$ and $J_0$ correspond to the non-zero and zero eigenvalues respectively, then
\begin{equation}
\label{eq54}
\bold{A}^\mathcal{D} = \bold{V}
\begin{pmatrix}
J_1^{-1} & 0\\
0 & 0\\
\end{pmatrix}
\bold{V}^{-1}
\end{equation}
It can be checked by direct substitution that this expression for $\bold{A}^\mathcal{D}$ satisfies Equations \ref{eq50}-\ref{eq52}.  We will use this to establish the following expression as the solution of Equation \ref{eq47}:  
\begin{equation}
\label{eq55}
\bold{x} = \bigg(-\bold{A}^\mathcal{D} + \bigg(\bold{I}-\bold{A}\bold{A}^\mathcal{D}\bigg)t\bigg[\sum_{n=0}^{k-1}\frac{\bold{A}^nt^n}{(n+1)!}\bigg] + e^{\bold{A}t}\bold{G}\bigg)\bold{b}
\end{equation}  

The following is a derivation of the results stated in \cite{campbell1976applications}.  First we show that
\begin{equation}
\label{eq56}
\int e^{-\bold{A}t} \text{dt} = -\bold{A}^\mathcal{D} e^{-\bold{A}t}+\bigg(\bold{I} - \bold{A}\bold{A}^\mathcal{D}\bigg) t \bigg[\sum_{n=0}^{k-1} \frac{(-1)^n\bold{A}^n t^n}{(n+1)!}\bigg]+\bold{G}
\end{equation}
where $\bold{G}$ is a constant of integration.  This can be done by taking the derivative of the right hand side and showing that it is equal to $e^{-\bold{A}t}$:

\begin{align*}
&\frac{\text{d}}{\text{dt}} \bigg(-\bold{A}^\mathcal{D}e^{-\bold{A}t} + \bigg(\bold{I}-\bold{A}\bold{A}^\mathcal{D}\bigg)t\bigg[\sum_{n=0}^{k-1} \frac{(-1)^n\bold{A}^n t^n}{(n+1)!}\bigg] +\bold{G}\bigg)\\
&=\bold{A}\bold{A}^\mathcal{D} e^{-\bold{A}t} +\bigg(\bold{I}-\bold{A}\bold{A}^\mathcal{D}\bigg)\bigg[\sum_{n=0}^{k-1} \frac{(-1)^n\bold{A}^n t^n}{(n+1)!}\bigg] + \bigg(\bold{I}-\bold{A}\bold{A}^\mathcal{D}\bigg)t\bigg[\sum_{n=0}^{k-1}\frac{(-1)^n n \bold{A}^n t^{n-1}}{(n+1)!}\bigg]
\end{align*}
Now we use the expansion for $e^{-\bold{A}t}$:  

\begin{align*}
\bold{A}\bold{A}^\mathcal{D}\bigg[\sum_{n=0}^{\infty}\frac{(-1)^n\bold{A}^nt^n}{n!}\bigg] + \bigg(\bold{I} - \bold{A}\bold{A}^\mathcal{D}\bigg) \bigg[\sum_{n=0}^{k-1}\frac{(-1)^n\bold{A}^nt^n}{(n+1)!}\bigg] + \bigg(\bold{I} -\bold{A}\bold{A}^\mathcal{D}\bigg)\bigg[\sum_{n=0}^{k-1}\frac{(-1)^nn \bold{A}^n t^n}{(n+1)!}\bigg]
\end{align*}
We combine the last two terms, canceling a factor of $(n+1)$, and then distribute the sum over $\bigg(\bold{I}-\bold{A}\bold{A}^\mathcal{D}\bigg)$:

\begin{align*}
\bold{A}\bold{A}^\mathcal{D} \bigg[\sum_{n=0}^{\infty}\frac{(-1)^n\bold{A}^nt^n}{n!}\bigg]+\sum_{n=0}^{k-1}\frac{(-1)^n\bold{A}^nt^n}{n!} - \bold{A}\bold{A}^\mathcal{D}\bigg[\sum_{n=0}^{k-1}\frac{(-1)^n\bold{A}^nt^n}{n!}\bigg]
\end{align*}
Subtracting the last term from the first term:

\begin{align*}
\bold{A}\bold{A}^\mathcal{D} \bigg[\sum_{n=k}^{\infty}\frac{(-1)^n\bold{A}^nt^n}{n!}\bigg]+\sum_{n=0}^{k-1}\frac{(-1)^n\bold{A}^nt^n}{n!} 
\end{align*}
Since $\bold{A}$ and $\bold{A}^\mathcal{D}$ commute we have:
\begin{align*}
\bold{A}^\mathcal{D} \bigg[\sum_{n=k}^{\infty}\frac{(-1)^n\bold{A}^{n+1}t^n}{n!}\bigg]+\sum_{n=0}^{k-1}\frac{(-1)^n\bold{A}^nt^n}{n!} 
\end{align*}
For $n\geq k$, $\bold{A}^\mathcal{D}\bold{A}^{n+1} = \bold{A}^n$ from Equation \ref{eq52}.  Using this we have:
\begin{align*}
&\sum_{n=k}^{\infty}\frac{(-1)^n\bold{A}^n t^n}{n!} + \sum_{n=0}^{k-1}\frac{(-1)^n\bold{A}^n t^n}{n!} 
=\sum_{n=0}^{\infty}\frac{(-1)^n\bold{A}^n t^n}{n!} \\
&=e^{-\bold{A}t}
\end{align*}
as desired.  Now that we have established Equation \ref{eq56}, we substitute it into Equation \ref{eq48}.  We assume here that $\bold{b}$ is constant.  

\begin{align*}
\bold{x} &= e^{\bold{A}t}\bigg(-\bold{A}^\mathcal{D}e^{-\bold{A}t} + \bigg(\bold{I} - \bold{A}\bold{A}^\mathcal{D}\bigg)t\bigg[\sum_{n=0}^{k-1}\frac{(-1)^n\bold{A}^n t^n}{(n+1)!}\bigg] +\bold{G}\bigg)\bold{b}\\
&=\bigg(-\bold{A}^\mathcal{D}+e^{\bold{A}t}\bigg(\bold{A}-\bold{A}\bold{A}^\mathcal{D}\bigg)t\bigg[\sum_{n=0}^{k-1}\frac{(-1)^n\bold{A}^nt^n}{(n+1)!}\bigg] + e^{\bold{A}t}\bold{G}\bigg)\bold{b}\\
&=\bigg(-\bold{A}^\mathcal{D}+\bigg(\bold{A}-\bold{A}\bold{A}^\mathcal{D}\bigg)t\bigg[\sum_{n=0}^{\infty}\frac{\bold{A}^nt^n}{n!}\bigg] \bigg[\sum_{n=0}^{k-1}\frac{(-1)^n\bold{A}^nt^n}{(n+1)!}\bigg] + e^{\bold{A}t}\bold{G}\bigg)\bold{b}
\end{align*}
It can be shown that the product of the two sums is 
\begin{equation*}
\sum_{n=0}^{k-1}\frac{\bold{A}^nt^n}{(n+1)!}
\end{equation*}
So we have

\begin{equation*}
\bold{x} = \bigg(-\bold{A}^\mathcal{D} + \bigg(\bold{I}-\bold{A}\bold{A}^\mathcal{D}\bigg)t\bigg[\sum_{n=0}^{k-1}\frac{\bold{A}^nt^n}{(n+1)!}\bigg] + e^{\bold{A}t}\bold{G}\bigg)\bold{b}
\end{equation*}
as claimed.  

It remains to establish the relationship between the constant of integration $\bold{G}$ and the initial condition $\bold{x}(0)$.  At $t=0$ we have

\begin{align*}
\bold{x}(0) = -\bold{A}^\mathcal{D}\bold{b} + \bold{G}\bold{b}
\end{align*}
Taking the transpose of both sides and rearranging we have
\begin{align*}
\bold{b}^\intercal\bold{G}^\intercal=\bold{x}(0)^\intercal + \bold{b}^\intercal{\bold{A}^{\mathcal{D}}}^{\intercal}
\end{align*}
Here ${\bold{A}^{\mathcal{D}}}^{\intercal}$ means the transpose of the Drazin inverse of $\bold{A}$.  We left multiply both sides by $\bold{b}$:
\begin{align*}
|\bold{b}|^2\bold{G}^\intercal = \bold{b}\bold{x}(0)^\intercal + |\bold{b}|^2 {\bold{A}^{\mathcal{D}}}^{\intercal}
\end{align*}
Dividing by $|\bold{b}|^2$, we have 
\begin{equation*}
\bold{G}^\intercal = \frac{\bold{b}}{|\bold{b}|^2}\bold{x}(0)^\intercal + {\bold{A}^{\mathcal{D}}}^{\intercal}
\end{equation*}
Taking the transpose of both sides, we have
\begin{equation}
\label{eq57}
\bold{G} = \frac{1}{|\bold{b}|^2}\bold{x}(0)\bold{b}^\intercal + \bold{A}^\mathcal{D}
\end{equation}
So the solution of $\bold{\dot{x}} = \bold{A}\bold{x} + \bold{b}$ is given by Equation \ref{eq55}, with the constant of integration chosen according to Equation \ref{eq57}.  We can put this all together to get 
\begin{equation}
\label{eq58}
\bold{x} = \bold{A}^\mathcal{D}\bold{b}\bigg(e^{\bold{A}t} - \bold{I}\bigg) + \bigg(\bold{I}-\bold{A}\bold{A}^\mathcal{D}\bigg)t\bigg[\sum_{n=0}^{k-1}\frac{\bold{A}^nt^n}{(n+1)!}\bigg]\bold{b} + e^{\bold{A}t}\bold{x}(0)
\end{equation}

\bibliographystyle{unsrt}